\documentclass{PoS}

\usepackage{graphicx}
\usepackage[latin1]{inputenc}
\usepackage{amssymb}
\usepackage{mathrsfs}

\usepackage{savesym}
\usepackage{amsmath}
\savesymbol{iint}
\usepackage[varg]{txfonts}
\restoresymbol{TXF}{iint}

\usepackage{amsfonts}
\usepackage{amssymb}
\usepackage{bm}
\usepackage{dsfont}
\usepackage{units}
\usepackage{cite}

\newcommand{\Slash}[1]{\ooalign{\hfil/\hfil\crcr$#1$}}

\title{On baryon properties from a covariant Faddeev approach}

\ShortTitle{On baryon properties from a covariant Faddeev approach}

\author{\speaker{Helios Sanchis-Alepuz},\\
       Institut f\"ur Physik, Karl-Franzens-Universit\"at Graz, Austria\\
       E-mail: \email{helios.sanchis-alepuz@uni-graz.at}}
\author{Reinhard Alkofer,\\
       Institut f\"ur Physik, Karl-Franzens-Universit\"at Graz, Austria}
\author{Gernot  Eichmann,\\
       Institut f\"ur Kernphysik, Technische Universit\"at Darmstadt, Germany}
\author{Selym  Villalba-Ch\'avez,\\
       Institut f\"ur Physik, Karl-Franzens-Universit\"at Graz, Austria}

\abstract{The Poincar\'e-covariant Faddeev equation and its solution in
rainbow-ladder truncation, i.e., with a generalized gluon exchange as
irreducible two-particle-interaction, is presented. The covariant decomposition
of baryon amplitudes, representing relativistic three-quark states, is
discussed and explicitly given for the $\Delta$-baryon. The calculated $\Delta$
mass compares favourably with the experimental one.}

\FullConference{Light Cone 2010: Relativistic Hadronic and Particle Physics\\
                 June 14-18, 2010\\
                 Valencia, Spain}

\begin{document}

\section{Introduction}

It is generally accepted that QCD is the theory of strong interactions and
thus hadrons are bound states of quarks and glue. Any numerical solution
of QCD's bound state equations, {\it i.e.}~the Bethe-Salpeter equation for
mesons and the Faddeev equation for baryons, requires Green functions of
quarks and gluons as input. To this end it is interesting to note that in  
Landau gauge the quark and gluon propagators are sufficiently well known for
phenomenological purposes; for recent reviews see, {\it e.g.\/},
\cite{Fischer:2006ub,Fischer:2008uz}. The aim is, of course, to use these
results, in particular the ones for the quark correlations, in the
relativistic bound state equations for baryons in a similar manner as for
mesons. However, the relativistic bound-state description for baryons is
considerably less understood than for mesons, even on the phenomenological
level. Therefore the complicated task of
implementing our knowledge of the strong-interaction domain of QCD in a
first-principle continuum description of baryons as Poincar{\'e}-covariant
three-quark bound states is only approached in small steps.

Over the last decade one has mostly followed an approach starting from a
covariant Faddeev equation \cite{Faddeev:1960su} but immediately reducing its
complexity by treating the two-body $T$-matrix in a separable expansion. Due
to the importance of quark-quark correlations, so-called diquarks, this
approximative treatment has been called quark-diquark approach
\cite{Hellstern:1997pg,Oettel:1998bk,Bloch:1999ke,Oettel:2000jj}. Reducing the Faddeev
equation to an effective two-body equation allowed for the calculation of many
different baryon observables; for a collection of recent results see, {\it
e.g.\/},
\cite{Eichmann:2007nn,Eichmann:2008ae,Nicmorus:2008vb,Nicmorus:2010sd,Eichmann:2010je}.

Only recently it was possible to solve the full  Poincar{\'e}-covariant
Faddeev equation
for the nucleon \cite{Eichmann:2009qa,Eichmann:2009zx,Eichmann:2009en}. Its
evolution with the pion mass reveals already quite interesting features,
especially when compared to corresponding lattice results. It still remains a
challenging task to compute observables other than the mass of the nucleon.

On the other hand, the lowest lying excited state of the nucleon, the
$\Delta(1232)$ resonance, enjoys a special place in the family  of baryon
resonances. The $\Delta (1232)$ resonance has the highest  production cross
section of all nucleon resonances, and it is a spin $3/2$  particle with the
same quark content as the nucleon. Therefore, it seems obvious to attempt also
for the $\Delta(1232)$ the corresponding solution of the Poincar{\'e}-covariant
Faddeev equation. In the following we will report on progress in this
direction by discussing the construction of the $\Delta$'s Faddeev amplitude
and by giving a (first) result for its mass.

\section{Faddeev equation}

Baryons in QCD appear as poles in the three-quark scattering matrix. Analyzing
the residue at the pole associated with a baryon of mass $M$ allows to derive
a relativistic equation for this bound state:
\begin{equation}
\boldsymbol{\Psi} =\pmb{K}_3\,\boldsymbol{\Psi}\,, \qquad \pmb{K}_3 =
 \pmb{K}_3^{irr} + \sum_{a=1}^3 \pmb{K}^{(a)}_{(2)}\,,
\end{equation}
where now the baryon is described on-shell by the bound-state amplitude
$\boldsymbol{\Psi}$. This equation includes all possible correlations among the
three quarks, which comprises a three-body irreducible contribution as well as
the sum of the three possible two-body irreducible interactions with a
spectator quark, denoted by $a$.

\begin{figure}[!htbp]
\begin{center}
\includegraphics[width=151mm]{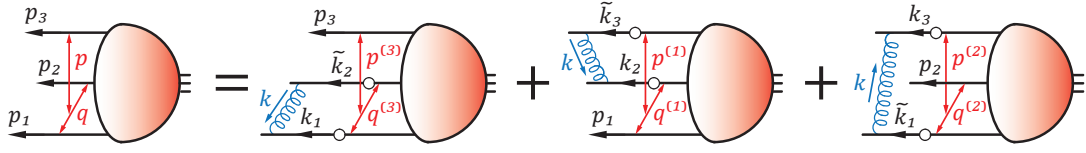}
\caption{Diagrammatical representation of  the Faddeev equation in rainbow-ladder truncation}
\end{center}
\end{figure}

The success of the quark-diquark approach to baryon properties supports the
idea that the quark-quark correlations dominate in binding baryons. Omitting
the three-body irreducible contribution leads to the covariant Faddeev
equation \cite{Faddeev:1960su}
\begin{equation}\label{EichmannG_faddeev:eq}
\boldsymbol{\Psi}_{\alpha\beta\gamma J}(p,q,P)  =\sum_{a=1}^3   \int\limits_k     \pmb{K}_{\alpha\alpha'\beta\beta'\gamma\gamma'}^{(a)} \, \boldsymbol{\Psi}_{\alpha'\beta'\gamma'J}(p^{(a)},q^{(a)},P)\,,
\end{equation}
where $\{\alpha,\beta,\gamma\}$ are Dirac indices for the three valence quarks
and the nature of the fourth index $J$ depends on the baryon of interest.
The momentum dependence of the amplitude on the three quark momenta $p_1$, $p_2$, $p_3$ 
is here reexpressed in terms of the total momentum $P$ and two relative momenta $p$ and $q$, 
where $P^2 = -M^2$ is fixed.
They are related via:
\begin{equation}
\begin{array}{ccc}
p = (1-\zeta)\,p_3 - \zeta p_d,\quad& q = \dfrac{p_2-p_1}{2},\quad&  P =
p_1+p_2+p_3,\\
~\\
p_1 =  -q -\dfrac{p}{2} + \dfrac{1-\zeta}{2} P,\quad &  p_2 =  q -\dfrac{p}{2} +
\dfrac{1-\zeta}{2} P,\quad&  p_3 =  p + \zeta  P ,
\end{array}
\end{equation}
where we abbreviated $p_d:=p_1+p_2$. We use the value $\zeta=1/3$ 
which maximizes the upper boundary for the nucleon mass with respect to restrictions arising
from the quark propagator's singularity structure; for a detailed description
see, {\it e.g.\/}, refs.\ \cite{Eichmann:2009zx,Oettel:2000ig}. 
The internal relative momenta are given by:
\begin{equation}
\begin{array}{ccc}
p^{(1)} = p+k,\quad& p^{(2)} = p-k,\quad& p^{(3)} = p,\\
~\\
q^{(1)} = q-k/2,\quad& q^{(2)} = q-k/2,\quad & q^{(3)} = q+k.
\end{array}
\end{equation}
Furthermore, $\pmb{K}^{(a)}$ denotes the renormalization-group invariant 
products of a $qq$ kernel and two dressed quark propagators:
\begin{eqnarray} \label{EichmannG_KSS}
\pmb{K}_{\alpha\alpha'\beta\beta'\gamma\gamma'}^{(1)}& =& \delta_{\alpha\alpha'}\mathcal{K}_{\beta\beta''\gamma\gamma''}\, S_{\beta''\beta'}(k_2)  \, S_{\gamma''\gamma'} (\widetilde{k}_3),\\
\pmb{K}_{\alpha\alpha'\beta\beta'\gamma\gamma'}^{(2)} &=& \delta_{\beta\beta'}\mathcal{K}_{\gamma\gamma''\alpha\alpha''}  S_{\gamma''\gamma'}(k_3)  S_{\alpha''\alpha'} (\widetilde{k}_1),\\
\pmb{K}_{\alpha\alpha'\beta\beta'\gamma\gamma'}^{(3)} &=& \delta_{\gamma\gamma'}\mathcal{K}_{\alpha\alpha''\beta\beta''}\, S_{\alpha''\alpha'}(k_1)  \, S_{\beta''\beta'} (\widetilde{k}_2).
\end{eqnarray} 
The quark propagators $S$  involved in these expressions depend on 
the internal quark momenta $k_i=p_i-k$ and $\widetilde{k}_j=p_j+k$, 
where $k$ is the gluon momentum.

\section{Rainbow-ladder truncation}

In order to perform our calculation we have to specify the $qq$ kernel as well
as the dressed propagators. We require our kernel to preserve the axial-vector
Ward-Takahashi identity, so that chiral symmetry and its dynamical breaking is
correctly implemented. The simplest realization of such a symmetry-preserving
truncation is  the rainbow-ladder truncation; for its definition within
relativistic three-body equations see refs.\ \cite{Taylor:1966zza,Eichmann:2009zx}. It 
describes the $qq$ kernel by a ladder dressed-gluon exchange
\begin{equation}\label{EichmannG_RL:kernel}
\mathcal{K}_{\alpha\alpha'\beta\beta'}(k) = Z_2^2 \,\frac{4\pi\alpha(k^2)}{k^2} \, T^{\mu\nu}_k \,\gamma^\mu_{\alpha\alpha'} \,\gamma^\nu_{\beta\beta'}
\end{equation}
which is  also involved  in the Dyson-Schwinger equation (for reviews see {\it
e.g.\/}, \cite{Fischer:2006ub,Roberts:2000aa,Alkofer:2000wg})
for the renormalized dressed quark propagator:
\begin{equation}\label{EichmannG_RL:QuarkDSE}
S^{-1}_{\alpha\beta}(p) = Z_2 \left( i\Slash{p} + m \right)_{\alpha\beta}  + \int_q \mathcal{K}_{\alpha\alpha'\beta'\beta}(k) \,S_{\alpha'\beta'}(q).
\end{equation}

Here the bare quark mass $m$ enters as an input and  $k=p-q.$ The color
structure  of the kernel generates  the prefactors $2/3$ and $4/3$ for the
integrals in Eq.\,(\ref{EichmannG_faddeev:eq}) and Eq.\,(\ref{EichmannG_RL:QuarkDSE}), respectively.
Eq.\,(\ref{EichmannG_RL:kernel}) involves an  effective coupling
$\alpha\left(k^2\right)$ for which we will henceforth use  the parametrization  
given  in~\cite{Maris:1999nt}:
\begin{equation}\label{EichmannG_couplingMT}
\alpha(k^2) = \pi \eta^7  \left(\frac{k^2}{\Lambda^2}\right)^2 \!\! e^{-\eta^2 \left(\frac{k^2}{\Lambda^2}\right)} + \alpha_{UV}\left(k^2\right).
\end{equation}
While  the first term describes the non-perturbative enhancement at small and
intermediate gluon momenta necessary to generate dynamical chiral symmetry
breaking and the dynamical generation of a constituent-quark mass scale, the
second term reproduces the one-loop QCD perturbative running coupling
$\alpha_{UV}(k^2)$ at large gluon momentum $(k^2\gg\Lambda_{QCD}^2).$   
The infrared behavior is parameterized by the dimensionless parameter
$\eta$ and the infrared scale $\Lambda$. Hereafter we  will fix the latter to
$\Lambda=0.72$~GeV  which has been used in
\cite{Eichmann:2007nn,Maris:1999nt,Krassnigg:2009zh} to reproduce experimental
results for meson and baryon properties. 
We remark that ground-state properties in this framework have turned out to be
insensitive to the choice of $\eta$ in a certain range \cite{Krassnigg:2009zh}.

\section{Covariant decomposition of  baryon amplitudes}

Numerical assessment  of the amplitude and mass of a particular  baryon  requires  to decompose  
the Faddeev amplitude in a suitable basis. This can be done independently of any approximation but based 
solely on first principles  like  Poincar\'e covariance, spatial and time-reversal symmetries 
and charge-conjugation properties. Considering this idea, a convenient basis for the nucleon amplitude  
was obtained in Refs.~\cite{Eichmann:2009zx,Eichmann:2009qa,Eichmann:2009en}. 
The latter turns out to be a four-rank tensor $\boldsymbol{\Psi}_{\alpha\beta\gamma\delta}$ 
(\textit{i.e.}, $J\equiv\delta$) in Dirac space  whose  dimension  is  $256$. 
However, for a nucleon of positive parity and positive energy its amplitude 
can be expanded in terms  of  a complete  orthogonal basis  which only involves  
$64$ Dirac structures:
\begin{equation}
\boldsymbol{\Psi}_{\alpha\beta\gamma\delta}=\sum_{\ell=1}^{64}\varkappa_\ell \,\boldsymbol{\Phi}_{\alpha\beta\gamma \delta}^{(\ell)}.\label{nucleondecomposition}
\end{equation}
Each  dressing function $\varkappa_\ell$ in Eq. (\ref{nucleondecomposition}) 
is a dimensionless  scalar factor which depends on five Lorentz-invariant combinations:
\[
p^2,\quad q^2,\quad z_0=\widehat{p_T}\cdot\widehat{q_T},\quad z_1=\hat{p}\cdot\hat{P},\quad z_2=\hat{q}\cdot\hat{P}.
\]
In our framework a  hat denotes a normalized four-vector\footnote{We work in Euclidean space.},
and $p^\mu_T=T^{\mu\nu}_P p^\nu$  with $T^{\mu\nu}_P=\delta^{\mu\nu}-\hat{P}^\mu\hat{P}^\nu$  
is a  transverse projector with respect to the baryon momentum.

It is noteworthy that  the basis $\left\{\boldsymbol{\Phi}\right\}$ can be divided 
into two sets $\left\{\mathcal{S}_{ij}^{r},\mathcal{P}_{ij}^{r}\right\}$ whose elements
\begin{equation}
\mathcal{S}_{ij}^r=\left(\mathbb{I}\otimes\mathbb{I}\right)\,\left(\Gamma_i\otimes \Gamma_j\right)\,\Omega_{r}(\hat{P})\qquad \mathrm{and} \qquad
\mathcal{P}_{ij}^r=\left(\gamma^5\otimes\gamma^5\right)\,\left(\Gamma_i\otimes \Gamma_j\right)\,\Omega_{r}(\hat{P})\,,\label{nucleonbasis1}
\end{equation}
with
\begin{equation}
\Omega_{r}(\hat{P})=\Lambda_r(\hat{P})\,\gamma^5\mathcal{C}\otimes \Lambda_+(\hat{P})\,,\label{nucleonbasis2}
\end{equation} 
involve  the charge conjugation-matrix $\mathcal{C}=\gamma^4\gamma^2$, 
the positive- and negative-energy projectors $\Lambda_\pm(\hat{P})=\left(\mathbb{I}\pm\Slash{\hat{P}}\right)/2$, 
and the elements encoded in
\begin{equation}
\Gamma_i=\left\{\,\mathbb{I},\;\frac{1}{2}\left[\widehat{\Slash{p_T}},\widehat{\Slash{q_t}}\right],\;\widehat{\Slash{p_T}},\;\widehat{\Slash{q_t}} \,\right\},
\end{equation}
where  $q_t^\mu=T^{\mu\lambda}_{p_T}T^{\lambda\nu}_P q^\nu=T^{\mu\lambda}_{p_T} q^{\lambda}_T$.  
Note that the  expression above introduces a relative-momen\-tum dependence in our amplitude. 
According to this notation, a particular basis element with index $\ell$ is fully determined by a combination  
of indices ${i,j, r}$ with $r=\pm$ and  $i,j=1,...,4.$

In contrast to the previous case, the covariant $\Delta-$amplitude is  a mixed tensor 
in  which five indices are involved. Four of them are spinorial  whereas the remaining one 
is  a four-vector index. As a consequence the  amplitude can be written as
\begin{equation}\label{eq:def-basis}
\boldsymbol{\Psi}_{\alpha\beta\gamma \delta}^{\mu}=\sum_{\ell}\varkappa_\ell \,\boldsymbol{\Phi}_{\alpha\beta\gamma \delta}^{(\ell)\,\mu}.
\end{equation}  
The spin-$3/2$ value of the $\Delta$ is then obtained  
by applying the  Rarita-Schwinger projector
\[
\mathbb{P}_+^{\mu\nu}(\hat{P})=\Lambda_+(\hat{P})\left(T^{\mu\nu}_P-\frac{1}{3}\gamma^\mu_T\gamma^\nu_T\right), \quad   \gamma^\mu_T=T^{\mu\nu}_P\gamma^\nu\,,
\]  
which is restricted to positive energy and obeys  the constrains 
$\hat{P}^\mu\,\mathbb{P}^{\mu\nu}_+(\hat{P})=\gamma_T^\mu\,\mathbb{P}^{\mu\nu}_+(\hat{P})=0.$

Now, in order to  construct the basis for Eq. (\ref{eq:def-basis}), we exploit  
the $64$ covariant structures given in  Eq.\,(\ref{nucleonbasis1}) and proceed
by multiplying  $\mathcal{S}_{ij}^{r}$ and $\mathcal{P}_{ij}^{r}$  
with all possible positive-parity combinations
\begin{eqnarray}\label{preliminacioin}
\begin{array}{ccccc}
\gamma^\mu_T\gamma^5\otimes \mathbb{I},\quad &   
\gamma^5\otimes\gamma^\mu_T, \quad &
\widehat{p_T}^\mu\gamma^5\otimes \mathbb{I},\quad &
\widehat{q_t}^\mu\gamma^5\otimes \mathbb{I},\quad & 
\hat{P}^\mu\gamma^5\otimes \mathbb{I},
\\ 
\mathbb{I}\otimes \gamma^\mu_T\gamma^5,\quad & 
\gamma_T^\mu\otimes\gamma^5, \quad &
\mathbb{I}\otimes \widehat{p_T}^\mu\gamma^5,\quad &
\mathbb{I}\otimes \widehat{q_t}^\mu\gamma^5,\quad &
\mathbb{I}\otimes \hat{P}^\mu\gamma^5.
\end{array}
\end{eqnarray}
This exercise  generates $640$ preliminary covariant structures which overcount the actual number of
independent basis elements. Indeed, because of the following identities
\begin{eqnarray}
\begin{array}{c}
\displaystyle
\left(\gamma^\mu_T\otimes\gamma^5 \right)
\left(\begin{array}{c}
\mathcal{S}_{ij}^r \\ \mathcal{P}_{ij}^r
\end{array}\right)=
\left(\gamma^\mu_T\gamma^5\otimes \mathbb{I}\right)
\left(\begin{array}{c}
\mathcal{P}_{ij}^r \\ \mathcal{S}_{ij}^r
\end{array}\right),\quad \left(\mathbb{I}\otimes\gamma^5 \right)
\left(\begin{array}{c}
\mathcal{S}_{ij}^r \\ \mathcal{P}_{ij}^r
\end{array}\right)=\left(\gamma^5\otimes\mathbb{I} \right)
\left(\begin{array}{c}
\mathcal{P}_{ij}^r \\ \mathcal{S}_{ij}^r
\end{array}\right),\\
~\\
\displaystyle
\left(\gamma^5\otimes \gamma^\mu_T\right)
\left(\begin{array}{c}
\mathcal{S}_{ij}^r \\ \mathcal{P}_{ij}^r
\end{array}\right)=\left(\mathbb{I}\otimes\gamma^\mu_T\gamma^5 \right)
\left(\begin{array}{c}
\mathcal{S}_{ij}^r \\ \mathcal{P}_{ij}^r
\end{array}\right)
\end{array}
\end{eqnarray} we can restrict ourselves to   the upper row in Eq.\,(\ref{preliminacioin}) and  halve the number given above  to $320.$  
A further reduction is achieved  by contracting the former
elements with $\mathbb{I}\otimes\mathbb{P}_+^{\mu\nu}$ which projects out the terms involving  $\hat{P}^\mu.$  
Consequently, $320-64=256$ covariant structures remain.  The  $64$ elements generated  from $\left(\gamma^5\otimes\gamma_T^\mu\right)$
can be ruled out as well since they turn out to be linear combinations of the remaining  $256-64=192$  basis elements:
\begin{center}
\begin{tabular}[h]{|c||c||c|}
\hline
$\left[\mathscr{G}_\mathfrak{g}\right]_{ij}^{r}$  & $\left[\mathscr{P}_\mathfrak{g}\right]_{ij}^{r}$ & $\left[\mathscr{Q}_\mathfrak{g}\right]_{ij}^{r}$
\\
\hline
\hline
&~&\\
 $\left(\gamma^\mu\gamma_5\otimes\mathbb{I}\right)\,\left(
\begin{array}{c}
\mathcal{S}_{ij}^r \\ \mathcal{P}_{ij}^r
\end{array}
\right)\,\left(\mathbb{I}\otimes\mathbb{P}_+^{\mu\nu}\right)$  &
 $\left(\gamma_5\otimes\mathbb{I}\right)\,\left(
\begin{array}{c}
\mathcal{S}_{ij}^r \\ \mathcal{P}_{ij}^r
\end{array}
\right)\,\left(\mathbb{I}\otimes \widehat{p_T}^\mu\mathbb{P}_+^{\mu\nu}\right)$ &
$\left(\gamma_5\otimes\mathbb{I}\right)\,\left(
\begin{array}{c}
\mathcal{S}_{ij}^r \\ \mathcal{P}_{ij}^r
\end{array}
\right)\,\left(\mathbb{I}\otimes \widehat{q_t}^\mu\mathbb{P}_+^{\mu\nu}\right)$ \\
&~&\\

\hline
\end{tabular}
\end{center}
in which  $\mathfrak{g}=\{\mathcal{S},\mathcal{P}\}$. Only $128$ of them  are linearly independent. 
We are free to choose a linearly independent subset; in particular, the following choice is orthonormal:
\begin{equation}\label{orthonormalbasis}
\begin{array}{c}
\sqrt{\frac{3}{2}}\left[\mathscr{P}_\mathfrak{g}\right]_{ij}^{r}\\
~\\
\sqrt{2}\left(\left[\mathscr{Q}_\mathfrak{g}\right]_{i1}^{r}-\frac{1}{2}\left[\mathscr{P}_\mathfrak{g}\right]_{i2}^{r}\right),\quad
\sqrt{2}\left(\left[\mathscr{Q}_\mathfrak{g}\right]_{i2}^{r}+\frac{1}{2}\left[\mathscr{P}_\mathfrak{g}\right]_{i1}^{r}\right),\\
~\\
\sqrt{2}\left(\left[\mathscr{Q}_\mathfrak{g}\right]_{i3}^{r}-\frac{1}{2}\left[\mathscr{P}_\mathfrak{g}\right]_{i4}^{r}\right),\quad
\sqrt{2}\left(\left[\mathscr{Q}_\mathfrak{g}\right]_{i4}^{r}+\frac{1}{2}\left[\mathscr{P}_\mathfrak{g}\right]_{i3}^{r}\right).
\end{array}
\end{equation}
We note that it is rather complicated to perform a partial-wave decomposition, {\it cf.}~\cite{Oettel:2000ig}, for this basis.
A better procedure would be to apply a Gram-Schmidt algorithm and include the
elements $\left[\mathscr{G}_\mathfrak{g}\right]_{11}^{r}$ which are likely
to be the dominant contributions. Such an analysis is beyond the  scope
of the present manuscript but a study in this direction  is in progress.

\section{Results and outlook}

Having defined an orthonormal basis for the $\Delta$-baryon amplitude, the
numerical method described in \cite{Eichmann:2009zx,Eichmann:2009en} can be
directly applied. The solution must have the correct symmetry under the
permutation group. Namely, both the flavour and the Poincar\'e-Dirac parts of
the amplitudes have to be fully symmetric under the exchange of any two quarks,
as required by the Pauli principle. Thanks to the iterative procedure used to
solve the equations, to fulfill this condition we just use
$\left[\mathscr{G}_\mathfrak{g}\right]_{11}^{+}$, which has the correct
symmetry, as the starting function. This element does not appear explicitly in
the basis~\eqref{orthonormalbasis} but can, of course, be expressed as a linear combination of the given basis
elements.

The resulting $\Delta$ mass at the physical pion mass is given in Table~1. We
see that the mass is slightly overestimated with respect to the physical value,
indicating that further structures need to be taken into account. Corrections
to this value, though small, would arise from considering kernels truncated beyond
rainbow-ladder and also including the $3$-body irreducible contributions in the
bound-state equation. On the other hand, the mass in this approach is only minimally below the
quark-diquark result~\cite{Nicmorus:2008vb} in which only the
axial-vector diquark was considered, which means that the diquark approximation works
already very well.

\begin{table}\label{result}
\begin{center}
\begin{tabular}[h]{|c||c||c|}
\hline
~&&\\
~~~~~$M_\Delta^{phys}$~~~~~ & ~~~~~$M_\Delta^{Q-DQ}$ ~~~~~
& ~~~~~$M_\Delta^{\rm Faddeev}$~~~~~\\
~&&\\
\hline
~&&\\
$1.23$ & $1.28$ & $1.26$\\
~&&\\
\hline
\end{tabular}
\end{center}
\caption{$\Delta$ mass (in GeV) obtained from the Faddeev equation  compared to the physical value and to the quark-diquark calculation.}
\end{table}

The results presented here are, however, the first piece in a more elaborated
investigation. First of all, as said above, it would be convenient to have a partial
wave decomposition of the basis to gain physical insight and ease comparison
with quark model calculations. Also, the dependence of our results with the
parameters in the effective coupling must be analyzed. Finally, the ingredients
needed to calculate the $\Delta$-baryon mass and structure allow us to perform
the same study for the $\Omega$-baryon, with the only difference being the
quark mass and the corresponding propagator. To compare with lattice data a
calculation for a range of quark masses is anyhow desirable.
Work in these directions is in
progress.

\acknowledgments

H.S.-A.\ thanks the organizers of this highly interesting workshop for the
possibility to present his recent work. We are grateful to M.\ Blank, C.S.\ Fischer, A.\ Krassnigg, D.\ Nicmorus
and R.\ Williams for helpful discussions. This work was supported by  the Austrian Science Fund FWF under Project No.\
P20592-N16, Erwin-Schr\"odinger Fellowship J3039-N16, and the Doctoral
Program W1203; as well as the Helmholtz Young Investigator Grant VH-NG-332 and in part by the European Union (HadronPhysics2
project ``Study of strongly-interacting matter'').

\end{document}